\documentstyle[twocolumn,prl,float,aps,epsfig,rotating]{revtex}
\begin{document}
\draft 

\title{Observation of double radiative capture on pionic hydrogen.}

\author{%
\mbox{ S.~Tripathi,$^{1}$} 
\mbox{ D.~S.~Armstrong,$^{4}$}  
\mbox{ M.~E.~Christy,$^{1}$} 
\mbox{ J.~H.~D.~Clark,$^{4}$\cite{APSaddress}} 
\mbox{ T.~P.~Gorringe,$^{1}$} 
\mbox{ M.~D.~Hasinoff,$^{2,3}$}
\mbox{ M.~A.~Kovash,$^{1}$} 
\mbox{ D.~H.~Wright,$^{3}$\cite{SLACaddress}}
\mbox{ P.~A.~{\.Z}o{\l}nierczuk,$^{1}$} 
\\
}

\address{%
\mbox{$^1$  University of Kentucky, Lexington, KY 40506}
\mbox{$^2$  University of British Columbia, Vancouver, B.C., Canada V6T 1Z1}
\mbox{$^3$  TRIUMF, 4004 Wesbrook Mall, Vancouver, B.C., Canada V6T 2A3} 
\mbox{$^4$  College of William and Mary, Williamsburg, VA 23187}
}

\date{\today}
\maketitle

\begin{abstract}
We report the first observation 
of double radiative capture on pionic hydrogen. 
The experiment was conducted at the TRIUMF cyclotron 
using the RMC spectrometer,
and detected $\gamma$--ray coincidences following $\pi^-$ stops
in liquid hydrogen.
We found the branching ratio for double radiative capture 
to be 
$( 3.05 \pm 0.27 (\mbox{stat.}) \pm 0.31  (\mbox{syst.}) ) \times 10^{-5}$.
The measured branching ratio and
angle-energy distributions support
the theoretical prediction 
of a dominant contribution from the $\pi \pi \rightarrow \gamma \gamma$ annihilation mechanism.
\end{abstract}



Negative pions stopped in hydrogen form pionic hydrogen atoms.
These atoms can disintegrate via several modes that include
the well-known processes of
charge exchange $\pi^- p \rightarrow \pi^o n$ \cite{Sp77}, 
radiative capture $\pi^- p \rightarrow \gamma n$ \cite{Sp77}, 
and pair production $\pi^- p \rightarrow e^+e^- n$ \cite{Sa61,Fo89}.


However, for pionic hydrogen an additional mode of capture 
is predicted by theory,
\begin{equation}
\pi^-p \rightarrow \gamma\gamma n \quad .
\end{equation}
This double-radiative process
has been investigated theoretically by several authors,
including Ericson and Wilkin \cite{Ericson:75},
Christillin and Ericson \cite{Christillin:79},
Gil and Oset \cite{Gil:95},
and Beder \cite{Beder:79}.
The predicted branching ratio is $5.1 \times 10^{-5}$ \cite{Beder:79},
with a mechanism
that is dominated
by the annihilation
of the  stopped, real $\pi^-$ 
on a soft, virtual $\pi^+$,
{\em i.e.} $\pi \pi \rightarrow \gamma \gamma$.
Beder also predicted different 
photon energy-angle distributions
for the contributing annihilation 
and bremsstrahlung mechanisms.


The underlying dynamics of $\pi\pi$ annihilation
in double radiative capture is rather intriguing.
For example, it led Ericson and Wilkin \cite{Ericson:75} 
to suggest the reaction
as a probe of the pion field in the nucleus,
and Gil and Oset \cite{Gil:95}
to suggest the reaction
as a novel window on the $\pi \pi \rightarrow \gamma \gamma$ vertex.
Also, the related $\gamma p \rightarrow \gamma \pi n$ reaction
was considered by Wolfe {\it et al.}~\cite{Wolfe:96} 
and Drechsel and Fil'kov~\cite{Drechsel:94}
as a possible probe of the pion polarizability.


The only experimental search for double radiative capture 
on pionic hydrogen was conducted 
by Vasilevsky {\it et al.}\ \cite{Vasilevsky:69} at JINR.
They used a large-acceptance photon-pair spectrometer 
and obtained a branching ratio upper limit of $5.5 \times 10^{-4}$.
However, double radiative capture on beryllium and carbon
has been observed in experiments 
by Deutsch {\it et al.}\ \cite{Deutsch:79} at CERN 
and Mazzucato {\it et al.}\ \cite{Mazzucato:80} at TRIUMF.
Unfortunately, these data are difficult to interpret
due to i) nuclear structure effects 
and ii) capture occurring from both the s- and p-states of
the $\pi$Be and $\pi$C atoms.



Our experiment was performed at the TRIUMF cyclotron using the 
RMC spectrometer \cite{Wright:92}.
The incident beam had
a pion flux of $7 \times 10^{5}$~s$^{-1}$,
a central momentum of 81.5~MeV/$c$,
and electron and muon contamination of 18\% and 9\% respectively.
The incoming pions 
were counted in a 4-element plastic scintillator telescope
and stopped in a 2.7 liter liquid hydrogen target
of length 15~cm, diameter 16~cm, and wall thickness 254~$\mu$m
\cite{Wright:92}.
The outgoing photons were detected 
by pair production 
in a 1~mm--thick cylindrical Pb converter 
and electron-positron tracking 
in cylindrical multiwire and drift chambers.  
A 1.2~kG axial magnetic field was used for momentum analysis
and concentric rings of segmented scintillators 
were used for fast-triggering.
The trigger scintillators comprised 
the A-ring (just inside the Pb converter radius), 
the C-ring (just inside the multiwire chamber radius),
and the D-ring (just outside the drift chamber radius).
For more information on the RMC spectrometer
see Wright {\it et al.}\ \cite{Wright:92}.
Note that in this experiment 
we moved the Pb converter 
from just inside the C-counter radius 
to just outside the A-counter radius.



\begin{figure}
\begin{center} 
\mbox{\epsfig{file=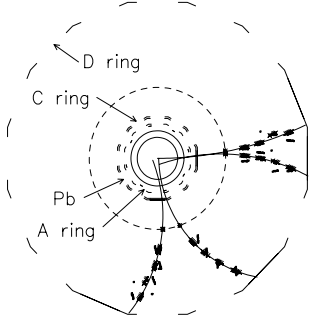,height=4.0cm,angle=90}}
\end{center}
\caption{A typical $\pi^- p \rightarrow \gamma \gamma n$ event.
The plot shows the fit in the plane perpendicular 
to the beam axis. 
The electron-positron pairs converge at the lead converter
and the reconstructed photon pairs originate from the hydrogen target located at the center.
The trigger pattern of
zero hits in the A-counter ring, 
two hits in the C-counter ring,
and four hits in the D-counter ring
is also displayed.
For scale, the radius of the D ring is about 60~cm.
}
\label{fig:event}
\end{figure}

For $\pi^-p\rightarrow\gamma\gamma n$ data-taking we employed
a two--photon trigger based
on the hit multiplicities 
and the hit topologies  
in the trigger scintillator rings 
and the drift chamber cells.
A typical $\pi^- p \rightarrow \gamma \gamma n$ event
that fulfilled the trigger is shown in Fig.\ \ref{fig:event}.
It has 
zero hits in the A-counter ring,
two hits in the C-counter ring,
and four hits in the D-counter ring.
To reduce the high rate 
of back-to-back photons 
from $\pi^o \rightarrow \gamma \gamma$ decay,
we rejected photon-pairs reconstructed with
drift cell hits or trigger scintillator hits
separated by large azimuthal angles.


During a four week running period
we collected $\pi^- p \rightarrow \gamma \gamma n$ data
from a total of $3.1 \times 10^{11}$ pion stops
in liquid hydrogen.
Calibration data 
with a dedicated $\pi^o \rightarrow \gamma \gamma$ trigger
were also taken periodically.


One source of background 
was real $\gamma$--$\gamma$ coincidences
arising from $\pi^o \rightarrow \gamma \gamma$ decay.
The $\pi^o$'s were produced by
either at-rest or in-flight pion charge exchange.
The at-rest source yields 
 $\pi^o$'s with energy $T = 2.8$~MeV 
and decay photons with opening angles $\cos{\theta} < -0.91$,
while the in-flight source yields 
$\pi^o$'s with $T \leq 15$~MeV
and photons with $\cos{\theta} < -0.76$.
The at-rest background was roughly $1600 \times$ 
the double-radiative capture signal,
and the in-flight background was roughly $10 \times$ 
the double radiative capture signal.
Consequently, for $\pi^-p\rightarrow\gamma\gamma n$
the $\cos{ \theta } < -0.76$ region
was overwhelmed by $\pi^o$ background,
and due to the finite resolution of the photon-pair spectrometer,
the $\pi^o$ background was a potential problem
for  opening angles with $\cos{\theta} > -0.76$.

Another source of background
was accidental $\gamma$--$\gamma$ coincidences
arising from simultaneous multiple $\pi^-$ stops.
The pion beam 
had a micro--structure with a pulse width of 2--4~ns
and a pulse separation of 43~ns.
With an incident flux of $7 \times 10^{5}$~s$^{-1}$
the probability for more than one pion arriving in a single beam pulse
is 1.5\%.
Multiple pion stops in one beam pulse
can yield a $\gamma$--ray pair
by the accidental coincidence of one photon
from each pion.
This background was roughly $900 \times$ our signal.
It yields photon-pairs with opening angles 0--180$^{\circ}$
and summed energies 106-258~MeV.
Note that the summed energy from random background events can exceed the m$_{\pi}$-limit for  single $\pi$ capture.


In analyzing the data a number of cuts were applied 
to identify photon pairs and reject background sources.
A tracking cut 
imposed minimum values for the number of points in the tracks
and maximum values for the chi--squared of fits to the tracks.
A photon cut 
required that the electron-positron pairs intersect at the Pb converter
and that the reconstructed photon pairs originate from the H$_2$ target.
To reject the multi-$\pi$ background we imposed a C-counter timing cut and
a beam telescope amplitude cut. 
The telescope amplitude cut was imposed 
on the normalized sum of the light output 
from the eight photo-multiplier tubes viewing
the four individual beam scintillators.
The $\pm$4~ns C-counter timing cut was imposed on
the time difference between the two C-counters intersecting
the two emerging e$^+$e$^-$ pairs.
The remaining inefficiency in rejecting 
the accidental coincidences was $1.3 \times 10^{-4}$.
To reject the $\pi^o$ background
we imposed a photon opening angle cut 
of $\cos{ \theta } > -0.1$.

\begin{figure}
\begin{center} 
\mbox{\epsfig{figure=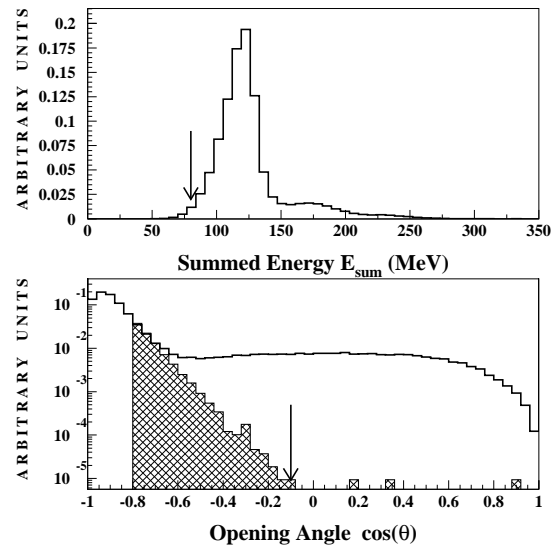,height=8.0cm}}
\end{center}
\caption{The summed energy spectrum (top) and opening angle spectrum (bottom)
for reconstructed photon-pairs, {\it i.e.} the events passing 
the tracking and photon cuts. 
Note that the multi--$\pi$ background can produce events 
with $E_{sum} > 150$~MeV (see top plot)
and the $\pi^o$ background will produce events 
with $\cos{ \theta } < -0.76$ (see bottom plot).
The Monte Carlo generated $\pi^o$ background is shown overlayed 
as the shaded histogram in the bottom plot. The arrow in the upper plot 
indicates the $E_{sum} > 80$~MeV cut and the arrow in the lower plot 
indicates the $\cos{ \theta } > -0.1$ cut.}
\label{fig:raw}
\end{figure}
\begin{figure}[ht]
\begin{center} 
\mbox{\epsfig{figure=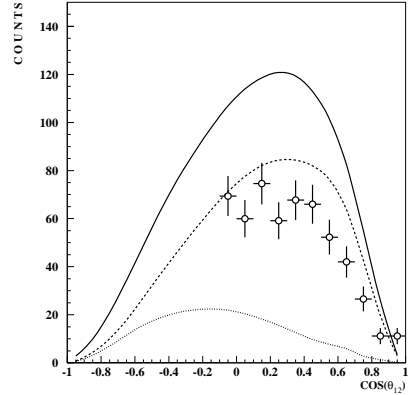,width=6cm}}
\end{center}
\caption{Comparison of the opening angle distributions
from the background subtracted experimental data (open circles)  and the theoretical calculation (curves).
The dashed curve is the $\pi\pi$ annihilation process,
the dotted curve is the  $NN$ bremsstrahlung process,
and the solid curve is the full calculation.
These curves are convoluted with the response function
of the RMC spectrometer.}
\label{fig:signal}
\end{figure}

A total of $2.3 \times 10^{6}$ photon pairs passed
both the tracking cuts and photon cuts.
These photon pairs
are shown in Fig.~\ref{fig:raw} and are dominated 
by the backgrounds from $\pi^o$ decays
and multi-$\pi$ stops. 
The multi-$\pi$ background
is clearly seen in the summed energy spectrum
as events with $E_{sum} > 150$~MeV
and the $\pi^o$ background
is clearly seen in the opening angle spectrum
as events with $\cos{ \theta } < -0.76$.
The beam telescope amplitude cut removed about $0.8 \times 10^{6}$
accidental $\gamma$--$\gamma$ coincidences from multi--$\pi$ stops,
and the photon opening angle cut removed about $1.4 \times 10^{6}$
real $\gamma$--$\gamma$ coincidences from $\pi^o$ decays.
A total of 635 events with 
$E_{sum} > 80$~MeV and $\cos{ \theta } > -0.1$ were found 
to survive all cuts (see Fig.~\ref{fig:signal}). 

A small quantity of two-photon background 
from $\pi^o$ decays and from multi--$\pi$ stops 
does however survive the applied cuts.
The remaining $\pi^o$ contamination was subtracted using 
(i) the observed number of $\pi^o$ events with $\cos{\theta} < -0.76$
and (ii) the known angular response of the photon-pair spectrometer.
The remaining multi--$\pi$ contamination was subtracted using 
(i) the observed number of 2$\pi$ events with $E_{sum} > 170$~MeV
and (ii) the measured sum energy spectrum for the 
multi-$\pi$ background.
These procedures indicated 53$\pm$30 $\pi^o$ background events, or 
($8.3 \pm 4.8$)\%,
and 100$\pm$16 multi-$\pi$ background events, 
or ($15.7 \pm 2.5$)\%,
with $E_{sum} > 80$~MeV
and $\cos{\theta} > -0.1$.
After subtraction this yielded a total 
of 482$\pm$42 $\pi^- p \rightarrow \gamma \gamma n$ events
with $E_{sum} > 80$~MeV 
and $\cos{\theta} > -0.1$.
Using the data from Refs.\ \cite{Deutsch:79,Mazzucato:80}
we further estimated the 
backgrounds originating from the 
nuclear $( \pi , 2 \gamma )$ reaction
on the target walls, etc,
to be $\leq 1$\%.


In order to compare the experiment with theory
we performed both measurements and simulations of the
two-photon response function of the RMC spectrometer.
To measure the response we employed the multi--$\pi$ accidentals.
Specifically, by counting the numbers of incoming multi-$\pi$ stops
and outgoing $\gamma$-$\gamma$ accidentals,
we mapped the detector's response
versus energy and angle\footnote{A minor complication is that the two photons
from the multi-$\pi$ stops
may have time differences
of up to 4~ns ({\it i.e} the pion beam pulse width).
This time difference slightly decreases
both the track reconstruction efficiency
and the two-photon acceptance.
From simulations we found the loss in acceptance 
to be 6\%.}.
Note that the calibration data and $\pi^- p\rightarrow \gamma \gamma n$ data
were collected simultaneously and passed through the same cuts 
(with the exception of the beam telescope cut).
Conveniently, the energy range and angular range
for multi-\-$\pi$ accidentals covers
the kinematical range for $\pi^- p \rightarrow \gamma \gamma n$ events.


To simulate the detector response function
we used a Monte Carlo program.
The program incorporated 
both the detailed geometry of the RMC detector
and the detailed interactions of the relevant particles.
Our program was based on the CERN GEANT3 package\cite{GEANT}
and is described in  Wright {\it et al.}\ \cite{Wright:92}.
We tested the simulation by comparison 
to the response function measurements
with the multi--$\pi$ accidentals.
We found the energy-angle distributions
from experiment and simulation
to be in very good agreement.
However the absolute detection efficiencies
from experiment 
and simulation were found to differ on average by 10\%.
Moreover, the measured acceptance was found
to vary by $\pm 4$\% from run to run
and decreased by 10\% between the early runs and the late runs.
We attributed these variations to
changes in the chamber efficiencies
and fluctuations in the chamber noise,
neither of which were incorporated in the simulation
of the acceptance.
To account for these differences between the measured
and the simulated acceptance, we employed a multiplicative correction factor 
of $F = 0.90 \pm 0.09$. The quoted uncertainty
is very conservative, and embodies
the  entire variation
of the measured acceptance
over the running period.


In addition we compared the results of
measurements and simulations of photon-pairs
from at-rest charge exchange $\pi^- p \rightarrow \pi^o n$ 
followed by neutral pion decay $\pi^o \rightarrow \gamma \gamma$.
Here we used dedicated `$\pi^0$ runs'
with a modified trigger arrangement
for these back-to-back photon pairs.
We found excellent agreement 
between the simulation and the measurements
using the factor $F = 0.90$.


The branching ratio for double radiative capture on pionic hydrogen
was obtained via
\begin{equation}
\label{e:br}
B.R. = \frac{N_{\gamma\gamma}}
{ N_{\pi^-} \cdot \epsilon\Omega \cdot F
\cdot c_{bm} \cdot c_{stop}} 
\end{equation}
where $N_{\pi^-}$ is the number 
of livetime-corrected pion stops,
$N_{\gamma \gamma}$ is the number 
of background-subtracted $\pi^- p \rightarrow \gamma \gamma n$ events,
and $\epsilon\Omega \cdot F$ is the detector acceptance.
Note that the appropriate acceptance was obtained 
using Monte Carlo~\cite{Wright:98} 
with the $\pi ^- p \rightarrow \gamma \gamma n$ kinematical distributions
taken from Beder \cite{Beder:79}.
The factor $c_{stop} = 0.85 \pm 0.01$ accounts
for the fraction of incident pions that stopped in hydrogen
(see Wright {\it et al.}~\cite{Wright:98} for details)
and the factor $c_{bm} = 0.99$ accounts
for the efficiency of $\pi^- p \rightarrow \gamma \gamma n$ events
passing the beam telescope cut.
Using Eqn.\ \ref{e:br}
we obtained a branching ratio of 
$( 3.05 \pm 0.27 (\mbox{stat.}) \pm 0.31  (\mbox{syst.}) ) \times 10^{-5}$. 
Note that the quoted uncertainty contains
a statistical error of $\pm 8$\% from $N_{\gamma \gamma}$ 
and a systematic error of $\pm 10$\% in total.
The systematic error is completely dominated by the $\pm 10$\% uncertainty 
in the determination of the acceptance $\epsilon\Omega \cdot F$.
The uncertainties in $N_{\pi^-}$, $c_{stop}$ and  $c_{bm}$
were each $\leq 2$ \% and entirely negligible.
We stress that the result we quote is the 
total $\pi^- p \rightarrow \gamma \gamma n$ branching ratio 
for all photon energies ($0 < E_{\gamma} < m_{\pi}$) 
and all opening angles ($-1.0 < \cos{ \theta } < +1.0$).

Our quoted branching ratio was extracted assuming 
the energy-angle distributions calculated by Beder \cite{Beder:79},
although we actually only observed the region
with $\cos{ \theta } > -0.1$ and $E_{\gamma} > 25$~MeV.
Tests of the sensitivity of the extracted branching ratio
to the energy-angle cuts revealed only a $\pm 2.9$\% 
variation for $-0.2 < \cos{ \theta } < 0.0$
and a $\pm 2.8$\% variation for analyses with different 
sum energy cuts.
Also, when using a phase space energy-angle distribution
rather than Beder's  energy-angle distribution~\cite{Beder:79}, 
the extracted branching ratio changed by only $-7$\%.


In Beder's calculation \cite{Beder:79} of double radiative capture 
the main contributions originate 
from $\pi\pi$ annihilation graphs, $NN$ bremsstrahlung graphs,
and their interference.
The $\pi\pi$ annihilation graphs alone account
for 64\% of the total branching ratio
and yield a distribution that is peaked at small opening angles.
The $NN$ bremsstrahlung graphs alone account
for 20\% of the total branching ratio
and yield a distribution that is peaked at large opening angles.
At threshold, the pion bremsstrahlung contributions vanish,
and effects of vector meson exchange and delta resonance excitation
are calculated to be very small.

In Fig.\ \ref{fig:signal} we compare our measured data 
with Beder's calculation \cite{Beder:79}.
The background contributions 
of ($8.3 \pm 4.8$)  \% from $\pi^o$ decay events
and ($15.7 \pm 2.5$)\% from multi-$\pi$ stop events have been subtracted 
from the measured data and the resulting 482 events are plotted as open 
circles.
The theoretical curves
have been convoluted 
with the response function of the RMC spectrometer.
The plot shows that
the $\pi^- p \rightarrow \gamma \gamma n$ branching ratio 
and opening angle distributions
from experiment and theory
are in reasonable agreement.
The overall consistency of experiment and theory 
supports  the theoretical prediction
of a dominant  $\pi\pi$ annihilation mechanism.
As seen in Fig.~\ref{fig:signal},
the bremsstrahlung graphs alone
underpredict the data
by about a factor of five.



However, our measured branching ratio
is somewhat smaller than the theoretical branching ratio. 
We note that Beder's calculation was performed at tree-level
and neglects contributions from pion loops, {\it etc}. 
We therefore speculate that higher--order terms might explain the 
difference
between the experimental  and the 
predicted branching ratio.
A new calculation 
of double radiative capture on pionic hydrogen 
using chiral perturbation theory is
currently underway \cite{Fearing:02}.


We remind the reader of the results from the previous nuclear $( \pi , 2 \gamma )$ measurements. For $^{12}$C, 
Deutsch {\it et al.}~\cite{Deutsch:79} obtained a partial branching ratio\ of 
$( 1.4 \pm 0.2 ) \times 10^{-5}$,
for $E_{\gamma} > 25$~MeV and $\cos{ \theta } < 0.71$,
and Mazzucato {\it et al.}~\cite{Mazzucato:80} obtained a partial branching ratio\ of 
$( 1.2 \pm 0.2 ) \times 10^{-5}$,
for $E_{\gamma} > 17$~MeV and $\cos{ \theta } < 0.71$.
Unfortunately the comparison of the earlier nuclear data
with our hydrogen data is difficult as
i) $\pi$ capture is predominantly from the 1S state in $^1$H
and the 2P state in $^{12}$C and 
ii) the nuclear data were mainly taken at large two-photon opening angles
and our hydrogen data were mainly taken  at small opening angles.


In summary, we have made the first measurement of  double radiative capture  
on pionic hydrogen by recording $\gamma$--ray coincidences 
from $\pi^-$ stops in liquid H$_2$.
We found the branching ratio 
to be  $( 3.05 \pm 0.27 (\mbox{stat.}) \pm 0.31  (\mbox{syst.}) ) \times 10^{-5}$
by assuming the kinematical distributions from Beder \cite{Beder:79}.
Moreover, the measured branching ratio and opening angle distribution 
support the theoretical hypothesis 
of a $\pi \pi$ annihilation mechanism.
We hope this work will stimulate further studies
into using double radiative capture as a novel probe of 
the proton's pion cloud and the 
$\pi \pi \rightarrow \gamma \gamma$ vertex.



We wish to thank the staff 
of the TRIUMF laboratory
for their support of this work.
In particular we acknowledge the help
of Ren\'{e}e Poutissou on the data acquisition
and Dennis Healey on the hydrogen target.
In addition we thank Douglas Beder 
and Harold Fearing for helpful discussions,
and the National Science Foundation (United States) 
the Natural Sciences and Engineering Research Council (Canada), 
the Henry Luce Foundation (JHDC) and the Jeffress Memorial Trust (DSA)  
for financial support.



\vspace{-0.5cm}

\begin{thebibliography}{99}

\vspace{-0.5cm}
\bibitem[\dag]{SLACaddress}Present address: 
SLAC, P.O. Box 20450, Stanford, CA 94309.

\bibitem[\ddag]{APSaddress}Present address:
American Physical Society, One Physics Ellipse, College Park, MD, 20740.

\bibitem{Sp77} J.~Spuller, D.~Berghofer, M.D.~Hasinoff, R.~Macdonald, 
D.F.~Measday, M.~Salomon, T.~Suzuki, J.M.~Poutissou, R.~Poutissou,
and J.K.P.~Lee, Phys.\ Lett.\ {\bf 67B}, 4 (1977). 

\bibitem{Sa61} N.P.~Samios, Phys.\ Rev.\ {\bf 121}, 275 (1961).

\bibitem{Fo89} H.~Fonvieille {\it et al.\ },
Phys.\ Lett.\ {\bf 233B}, 60 (1989). 

\bibitem{Ericson:75} T.E.O.~Ericson and C.~Wilkin, 
Phys.\ Lett.\ {\bf 57B}, 345 (1975).

\bibitem{Christillin:79} P.~Christillin and T.E.O.~Ericson, 
Phys.\ Lett.\ {\bf 87B}, 163 (1979).

\bibitem{Gil:95} A.~Gil and E.~Oset, Phys.\ Lett.\ {\bf 346B}, 1 (1995).

\bibitem{Beder:79} D.~Beder, Nucl.\ Phys.\ {\bf B156}, 482 (1979).

\bibitem{Nyman:77}  Ebbe M. Nyman and Mannque Rho,
Nucl.\ Phys.\ {\bf A287}, 390-398 (1977).

\bibitem{Wolfe:96} C.E.~Wolfe, S.~Nozawa, M.N.~Butler and B.~Castel,
Int.\ Jour.\ of Mod.\ Phys.\, {\bf E5}, 227 (1996).

\bibitem{Drechsel:94}
D.~Drechsel and L.~V.~Fil'kov,
Z.\ Phys.\ A {\bf 349}, 177 (1994).

\bibitem{Vasilevsky:69} I.M.~Vasilevsky, V.V.~Vishnyakov, A.F.~Dunaytsev, 
Yu.D.~Prokoshkin, V.I.~Rykalin, and A.A.~Tyapkin, 
Nucl.\ Phys.\ {\bf B9}, 673 (1969).

\bibitem{Deutsch:79} J.~Deutsch, D.~Favart, M.~Lebrun, P.~Lipnik, P.~Macq
and R.~Prieels, Phys.\ Lett.\ {\bf 80B}, 347 (1979).

\bibitem{Mazzucato:80} M. Mazzucato, B. Bassaleck, M.D. Hasinoff, T. Marks,
J.M. Poutissou and M. Salomon,  Phys.\ Lett.\ {\bf 96B}, 43 (1980).

\bibitem{Wright:92} D.H.~Wright {\em et~al.}, 
Nucl.\ Instrum.\ Methods {\bf A320}, 249 (1992).

\bibitem{Wright:98} D.H.~Wright {\em et~al.}, Phys.\ Rev.\ C {\bf 57}, 373 (1998).

\bibitem{GEANT} R.~Brun, F.~Bruyant, M.~Maire, A.C.~McPherson and
P.~Zanarini, GEANT3(1986); CERN report no. DD/EE/84--1 (unpublished).

\bibitem{Fearing:02} H.W.~Fearing, private communication.

\end{thebibliography}
\end{document}